\documentstyle [12pt]{article}
\begin{document}
\newtheorem{prop}{Proposition}
\newtheorem{cor}[prop]{Corollary}
\renewcommand {\theequation}{\thesection.\arabic{equation}}
\newcommand {\beq}{\begin{equation}}
\newcommand {\eeq}{\end{equation}}
\newcommand {\beqa}{\begin{eqnarray}}
\newcommand {\eeqa}{\end{eqnarray}}
\newcommand {\sfrac}[2]{{\textstyle \frac{#1}{#2}}}
\newcommand {\n}{\nonumber \\}
\newcommand {\eqn}[1]{(\ref{#1})}
\newcommand {\eq}[1]{eq.(\ref{#1})}
\newcommand {\eqs}[1]{eqs.(\ref{#1})}
\newcommand {\Eq}[1]{Eq.(\ref{#1})}
\newcommand {\Eqs}[1]{Eqs.(\ref{#1})}
\newcommand {\Label}[1]{\label{#1}}
\newcommand {\Bibitem}[1]{\bibitem{#1}}
\newcommand {\eqdef}{\stackrel{\rm def}{=}}
%
\font\tennn=msbm10
\font\twelvenn=msbm10 scaled\magstep1
%
\newcommand {\Sbm}[1]{\leavevmode\raise-.25ex\hbox{\twelvenn #1}}
\newcommand {\sbm}[1]{\leavevmode\raise-.15ex\hbox{\tennn #1}}
\begin{flushright}
{\boldmath $Y\!ukawa$ $Institute$ $Kyoto$}\hfill
RIMS-924\\
YITP/K-1018\\
May 1993\\
\end{flushright}
\vspace{15mm}
\begin{center}
{\LARGE Free Boson Realization of $U_q(\widehat{sl_N})$}\\
\vspace{25mm}
{\large
 Hidetoshi Awata\footnote{Fellow of Soryushi Shogakukai.},
 Satoru Odake${}^{1,}$\footnote{Address after June 1;
  Department of Physics, Faculty of Liberal Arts,
  Shinshu University, Matsumoto 390, Japan.}
 }\\
\vspace{5mm}
{\large {\sl Yukawa Institute for Theoretical Physics\\
              Kyoto University, Kyoto 606-01, Japan}}\\
\vspace{5mm}
and\\
\vspace{5mm}
{\large
 Jun'ichi Shiraishi\footnote{
 On leave from Department of Physics, University of Tokyo,
 Tokyo 113, Japan.}}\\
\vspace{5mm}
{\large {\sl Research Institute for Mathematical Sciences\\
              Kyoto University, Kyoto 606-01, Japan}}
\end{center}
\vspace{20mm}
\begin{abstract}
We construct a realization of the quantum affine algebra
$U_q(\widehat{sl_N})$ of an arbitrary level $k$ in terms of free
boson fields. In the $q\!\rightarrow\! 1$ limit this realization
becomes the Wakimoto realization of $\widehat{sl_N}$.
The screening currents and the vertex operators(primary fields)
are also constructed; the former commutes with $U_q(\widehat{sl_N})$
modulo total difference, and the latter creates the
$U_q(\widehat{sl_N})$ highest weight state from the vacuum state
of the boson Fock space.
\end{abstract}
\newpage
%
%
\section{Introduction}
\setcounter{equation}{0}
Chiral algebras such as the Virasoro and current algebras play
a central role in the conformal field theory (CFT) in two dimensional
space-time. This theory is a quantum field theory (QFT) of massless
particles, in other words, a (massive) QFT at a critical point
(renormalization-group fixed point)\cite{BPZ}.
Perturbing CFT's suitably, we get integrable massive QFT's
\cite{Za,EY,FF}. In these theories, the Virasoro algebra does
not exist any longer. In many cases the quantum affine Lie
algebra plays a crucial role instead of the Virasoro algebra\cite{BL}.
This quantum algebra is, for a large part, at the origin of the
integrability. Moreover it can almost determine the S-matrix of the
theory, e.g. sine-Gordon model\cite{BL}.

The Wess\,-Zumino\,-Novikov\,-Witten (WZNW) model is a fundamental
example of CFT's;
many CFT's can be realized through a coset construction
of WZNW models. The WZNW model has been studied based on the
representation theory of the affine Lie algebra.
Correlation functions of this model, which are vacuum
expectation values of vertex operators, satisfy certain holomorphic
differential equations, what is called, Knizhnik-Zamolodchikov (KZ)
equations\cite{KZ,TK}.
We expect that \lq\lq$q$-WZNW model", which has a symmetry of the
quantum affine algebra, is certain massive deformation of the WZNW
model. Correlation functions of the $q$-WZNW model satisfy
$q$-difference equations ($q$-KZ equations)\cite{Sm,FR}.
Connection matrices of solutions for $q$-KZ equations are related
to elliptic solutions of the Yang-Baxter equations of RSOS models
\cite{FR}.
An application of $q$-vertex operators based on $U_q(\widehat{sl_2})$
was performed in diagonalization of the XXZ spin chain\cite{DFJMN}.

Free field realizations of the Virasoro and affine Lie algebras
were useful for studying representation theories\cite{FFu} and
calculating correlation functions\cite{DF,ATY}.
It is expected that this is also the case for the quantum affine
algebras. In fact, the integral formula for correlation
functions of the local operators of the XXZ spin chain was found by
using the free boson realization of $U_q(\widehat{sl_2})$ and bosonized
$q$-vertex operators\cite{JMMN,IIJMNT}. To study higher rank versions
of the XXZ spin chain, sine-Gordon model, etc., we need free field
realizations of the quantum affine algebras.

In this paper we construct a free boson realization of the quantum
affine algebra $U_q(\widehat{sl_N})$ with an arbitrary level $k$.
In the $q\rightarrow 1$ limit, it becomes the bosonized version of
the Wakimoto realization of $\widehat{sl_N}$\cite{W,FF2,IKKS}.
Free field realizations of $U_q(\widehat{sl_N})$ with
level $1$ were constructed in \cite{FJ}.
Free field realizations of $U_q(\widehat{sl_2})$ with an arbitrary
level were constructed by several authors\cite{S,M,ABG,KB} and
that of $U_q(\widehat{sl_3})$ was obtained by the present authors
\cite{AOS}.
We construct a free boson realization of $U_q(\widehat{sl_N})$
by affinizing the Heisenberg realization ($q$-difference
operator realization) of $U_q(sl_N)$\cite{ANO}
and prove it by the OPE (operator product expansion) technique.
The screening currents and the vertex operators(primary fields) are
also constructed. They are necessary ingredients for calculating
correlation functions. Certain integral of the screening current
commutes with $U_q(\widehat{sl_N})$ and the vertex operator creates
the highest weight state of $U_q(\widehat{sl_N})$ from the vacuum
state of the boson Fock space.

This paper is organized as follows. In section 2 we fix our notations
and recall the definition of $U_q(\widehat{sl_N})$. We construct
a free field realization of $U_q(\widehat{sl_N})$ in section 3,
and the screening currents and the vertex operators in section 4.
Section 5 is devoted to discussion. The grading operator is also
bosonized.
In Appendix A we present the Heisenberg realization of $U_q(sl_N)$.
In Appendix B $q$-difference expressions of our free field realization
are given. In Appendix C,D we give useful formulas and some details of
calculations.

%
%
%
%
%
%
\section{Notations}
\setcounter{equation}{0}
Throughout this paper, complex numbers $q$ and $k$ are fixed.
$q$ is assumed to be a generic value with $|q|<1$.
We will use the standard symbol $[x]$,
\beq
 [x]\eqdef \frac{q^x-q^{-x}}{q-q^{-1}},
\eeq
and $\sum_{r=n}^{n-1}* \eqdef 0$, $\prod_{r=n}^{n-1}* \eqdef 1$.
Let $\bar{\alpha}_i$, $\bar{\Lambda}_i$ $(1\leq i\leq N-1)$,
$(a_{ij})_{1\leq i,j \leq N-1}$, be the simple roots, fundamental
weights, the Cartan matrix of $sl_N$ respectively.
$(\cdot,\cdot)$ is the symmetric bilinear form;
$(\bar{\alpha}_i,\bar{\alpha}_j)=a_{ij}$,
$(\bar{\Lambda}_i,\bar{\alpha}_j)=\delta_{ij}$.
$g$ stands for the dual Coxeter number of $sl_N$, i.e., $g=N$.

The $q$-difference operator with a parameter $\alpha$ is defined
by\cite{S}
\beq
 {}_{\alpha}\partial_z f(z) \eqdef
 \frac{f(q^{\alpha}z)-f(q^{-\alpha}z)}{(q-q^{-1})z}.
 \Label{qdiff}
\eeq
The Jackson integral with parameters $p\in\Sbm{C}$ $(|p|<1)$ and
$s\in\Sbm{C}^{\times}$ is defined by
\beq
 \int_0^{s\infty}f(z)d_pz \eqdef
 s(1-p)\sum_{n\in \sbm{Z}} f(sp^n)p^n.
 \Label{Jint}
\eeq
These operations satisfy the following property:
\beq
 \int_0^{s\infty}{}_{\alpha}\partial_z f(z) d_pz =0~~~
 \mbox{for }p=q^{2\alpha}.
 \Label{Jint0}
\eeq
The deformed commutator with a parameter $p\in\Sbm{C}$ is
\beq
 \lbrack A,B\rbrack_p \eqdef AB-pBA.
\eeq

The quantum affine algebra $U_q(\widehat{sl_N})$ is the associative
algebra over \Sbm{C} with Chevalley generators $e^{\pm}_i$,
invertible $t_i$ $(i=0,1,\cdots,N-1)$, and the following
relations\cite{JD}\footnote{For the grading operator $d$,
 see section 5.}:
\beqa
 \lbrack t_i,t_j \rbrack &\!\!=\!\!& 0, \\
 t_i e^{\pm}_j t_i^{-1} &\!\!=\!\!& q^{\pm a^{ext}_{ij}} e^{\pm}_j, \\
 \lbrack e^+_i,e^-_j \rbrack &\!\!=\!\!&
  \delta_{ij}\frac{t_i-t_i^{-1}}{q-q^{-1}},
\eeqa
and
\beq
 \sum_{r=0}^{1-a^{ext}_{ij}} (-1)^r
 \biggl\lbrack {1-a^{ext}_{ij}\atop r} \biggr\rbrack
 (e^{\pm}_i)^{1-a^{ext}_{ij}-r} e^{\pm}_j (e^{\pm}_i)^r =0,
\eeq
where $(a^{ext}_{ij})_{0\leq i,j\leq N-1}$ is the Cartan matrix of
the extended Dynkin diagram of $sl_N$ and
$[{n\atop r}]\eqdef\frac{[n]!}{[r]![n-r]!}$,
$[n]! \eqdef \prod_{r=1}^n [r]$.

$U_q(\widehat{sl_N})$ is isomorphic to the associative algebra
over \Sbm{C} with Drinfeld generators $E^{\pm,i}_n$ $(n\in\Sbm{Z})$,
$H^i_n$ $(n\in\Sbm{Z}-\{ 0\})$, invertible $K_i$
$(i=1,2,\cdots,N-1)$, invertible $\gamma$,
and the following relations\cite{D}:
\beqa
 \gamma &:& \mbox{central element},
 \Label{gamma} \\
 \lbrack K_i, H^j_n \rbrack &\!\!=\!\!& 0,~~~
 K_i E^{\pm,j}_n K_i^{-1}=q^{\pm a_{ij}}E^{\pm,j}_n,
 \Label{KHE} \\
 \lbrack H^i_n, H^j_m \rbrack &\!\!=\!\!&
 \frac{1}{n}[a_{ij}n]
 \frac{\gamma^n-\gamma^{-n}}{q-q^{-1}}\delta_{n+m,0},
 \Label{HnHm} \\
 \lbrack H^i_n, E^{\pm,j}_m \rbrack &\!\!=\!\!&
 \pm\frac{1}{n}[a_{ij}n]\gamma^{\mp\frac{1}{2}|n|}E^{\pm,j}_{n+m},
 \Label{HnEm} \\
 \lbrack E^{+,i}_n, E^{-,j}_m \rbrack &\!\!=\!\!&
 \frac{\delta^{ij}}{q-q^{-1}}
 \Bigl( \gamma^{\frac{1}{2}(n-m)}\psi^i_{+,n+m}
       -\gamma^{-\frac{1}{2}(n-m)}\psi^i_{-,n+m} \Bigr),
\eeqa
and
\beqa
 &&
 \lbrack E^{\pm,i}_{n+1}, E^{\pm,j}_m \rbrack_{q^{\pm a_{ij}}}
 +\lbrack E^{\pm,j}_{m+1}, E^{\pm,i}_n \rbrack_{q^{\pm a_{ij}}}=0,\\
 &&
 \lbrack E^{\pm,i}_n, E^{\pm,j}_m \rbrack=0 ~~~\mbox{for } a_{ij}=0,
 \Label{EE0mode} \\
 &&
 \lbrack E^{\pm,i}_n, \lbrack E^{\pm,i}_m, E^{\pm,j}_{\ell}
  \rbrack_{q^{\mp 1}} \rbrack_{q^{\pm 1}}
 +\lbrack E^{\pm,i}_m, \lbrack E^{\pm,i}_n, E^{\pm,j}_{\ell}
  \rbrack_{q^{\mp 1}} \rbrack_{q^{\pm 1}} =0 ~~~\mbox{for } a_{ij}=-1.
 \Label{EEEmode}
\eeqa
Here $\psi^i_{\pm,n}$ are defined by the following equation:
\beq
 \sum_{n\in\sbm{Z}} \psi^i_{\pm,n} z^{-n}
 \eqdef K_i^{\pm 1}
 \exp\Bigl( \pm (q-q^{-1})\sum_{\pm n>0} H^i_n z^{-n} \Bigr).
 \Label{psin}
\eeq
Let $H^i_0$ be defined by
\beq
 K_i\eqdef\exp((q-q^{-1})\sfrac{1}{2}H^i_0),
 \Label{K}
\eeq
then \eqs{KHE}-\eqn{HnEm} hold for $H^i_n$ $(n\in\Sbm{Z})$\footnote{
 In the case of $n=0$, $\frac{1}{n}*$ should be understood as
 $\lim_{n\rightarrow 0}\frac{1}{n}*$. For example,
 $\lim_{n\rightarrow 0}\frac{1}{n}[n]=\frac{2\log q}{q-q^{-1}}$,
 $\lim_{n\rightarrow 0}\frac{1}{n}[a_{ij}n]
  \frac{\gamma^n-\gamma^{-n}}{q-q^{-1}}=0$,
 $\lim_{n\rightarrow 0}\frac{1}{n}[a_{ij}n]
  \gamma^{\mp\frac{1}{2}|n|}=\frac{2\log q}{q-q^{-1}}a_{ij}$.
 In the following, this convention is assumed.
}. \Eq{KHE} is derived from \eqs{HnHm},\eqn{HnEm}.

Defining the fields $H^i(z)$, $E^{\pm,i}(z)$ and $\psi^i_{\pm}(z)$ as
\beq
 H^i(z) \eqdef \sum_{n\in \sbm{Z}}H^i_n z^{-n-1},~~~
 E^{\pm,i}(z) \eqdef \sum_{n\in \sbm{Z}}E^{\pm,i}_n z^{-n-1},~~~
 \psi^i_{\pm}(z) \eqdef \sum_{n\in \sbm{Z}}\psi^i_{\pm,n} z^{-n},
\eeq
the above relations can be rewritten as formal power series equations:
\beqa
 &&
 \lbrack\psi^i_{\pm}(z),\psi^j_{\pm}(w)\rbrack
 = 0, \\
 &&
 (z-q^{a_{ij}}\gamma^{-1}w)(z-q^{-a_{ij}}\gamma w)
 \psi^i_+(z) \psi^j_-(w) \n
 &&
 \hspace{5mm}=
 (z-q^{a_{ij}}\gamma w)(z-q^{-a_{ij}}\gamma^{-1}w)
 \psi^j_-(w) \psi^i_+(z),
 \Label{psizpsiw} \\
 &&
 (z-q^{\pm a_{ij}}\gamma^{\mp\frac{1}{2}}w)\psi^i_+(z) E^{\pm,j}(w)
 =
 (q^{\pm a_{ij}}z-\gamma^{\mp\frac{1}{2}}w)E^{\pm,j}(w) \psi^i_+(z),\\
 &&
 (z-q^{\pm a_{ij}}\gamma^{\mp\frac{1}{2}}w)E^{\pm,j}(z) \psi^i_-(w)
 =
 (q^{\pm a_{ij}}z-\gamma^{\mp\frac{1}{2}}w)\psi^i_-(w) E^{\pm,j}(z),\\
 &&
 \lbrack E^{+,i}(z),E^{-,j}(w)\rbrack
 =
 \frac{\delta^{ij}}{(q-q^{-1})zw}
 \Bigl(
 \delta(z^{-1}w\gamma)\psi^i_+(\gamma^{\frac{1}{2}}w)
 -\delta(z^{-1}w\gamma^{-1})\psi^i_-(\gamma^{-\frac{1}{2}}w)
 \Bigr),~~~~~~
 \Label{E+E-}
\eeqa
and
\beqa
 &&
 (z-q^{\pm a_{ij}}w)E^{\pm,i}(z)E^{\pm,j}(w)=
 (q^{\pm a_{ij}}z-w)E^{\pm,j}(w)E^{\pm,i}(z),
 \Label{EzEw} \\
 &&
 E^{\pm,i}(z)E^{\pm,j}(w) = E^{\pm,j}(w)E^{\pm,i}(z)~~~
 \mbox{for } a_{ij}=0,
 \Label{EE0} \\
 &&
 E^{\pm,i}(z_1)E^{\pm,i}(z_2)E^{\pm,j}(w)
 -(q+q^{-1}) E^{\pm,i}(z_1)E^{\pm,j}(w)E^{\pm,i}(z_2) \n
 &&
 \hspace{5mm}
 +E^{\pm,j}(w)E^{\pm,i}(z_1)E^{\pm,i}(z_2)
 +(\mbox{replacement}:z_1\leftrightarrow z_2)= 0~~~
 \mbox{for } a_{ij}=-1,
 \Label{EEE}
\eeqa
where $\delta(x)$ is given by
\beq
 \delta(x)\eqdef\sum_{n\in\sbm{Z}}x^n.
\eeq

Correspondence between Chevalley generators and Drinfeld generators
are \cite{D}:
\beqa
 t_i &\mapsto& K_i~~~(i=1,\cdots,N-1), \\
 e^{\pm}_i &\mapsto& E^{\pm,i}_0~~~(i=1,\cdots,N-1), \\
 t_0 &\mapsto& \gamma K_1^{-1}\cdots K_{N-1}^{-1}, \\
 e^+_0 &\mapsto&
 \lbrack E^{-,N-1}_0,
 \lbrack E^{-,N-2}_0,
 \lbrack \cdots,
 \lbrack E^{-,2}_0, E^{-,1}_1
 \rbrack_{q^{-1}} \cdots \rbrack_{q^{-1}} \rbrack_{q^{-1}}
 K_1^{-1}\cdots K_{N-1}^{-1}, \\
 e^-_0 &\mapsto&
 K_1\cdots K_{N-1}
 \lbrack \lbrack \cdots \lbrack
 E^{+,1}_{-1},E^{+,2}_0
 \rbrack_q, \cdots,
 E^{+,N-2}_0 \rbrack_q, E^{+,N-1}_0
 \rbrack_q.
\eeqa
$U_q(\widehat{sl_N})$ has the Hopf algebra structure.
We take its coproduct $\Delta$ as
\beqa
 \Delta (t_i) &\!\!=\!\!& t_i\otimes t_i, \\
 \Delta (e^+_i) &\!\!=\!\!& e^+_i\otimes 1+t_i\otimes e^+_i, \\
 \Delta (e^-_i) &\!\!=\!\!& e^-_i\otimes t_i^{-1}+1\otimes e^-_i,
\eeqa
and its antipode $S$ is
\beq
 S(t_i)=t_i^{-1},~~~S(e^+_i)=-t_i^{-1}e^+_i,~~~
 S(e^-_i)=-e^-_it_i.
\eeq
An explicit coproduct formula for all the Drinfeld generators
has not been obtained.

Let $V(\lambda)$ be the Verma module over $U_q(\widehat{sl_N})$
generated by the highest weight state $|\lambda\rangle$, such that
\beqa
 H^i_n|\lambda\rangle &\!\!=\!\!&
 E^{\pm,i}_n|\lambda\rangle=0~~~(n>0), \\
 E^{+,i}_0|\lambda\rangle &\!\!=\!\!& 0, \\
 H^i_0|\lambda\rangle &\!\!=\!\!& \ell^i|\lambda\rangle,
\eeqa
where the classical part of the highest weight is
$\bar{\lambda}=\sum_{i=1}^{N-1}\ell^i\bar{\Lambda}_i$.

Next we will introduce boson fields.
For a set of bosonic oscillators $a_n$ ($n\in \Sbm{Z}$),
and zero modes $\hat{p}_a$, $\hat{q}_a$ whose commutation relations are
\beqa
 \lbrack a_n,a_m \rbrack &\!\!=\!\!& n\rho_a(n) \delta_{n+m,0},
 \hspace{5mm}
 a_0=\frac{2\log q}{q-q^{-1}} \hat{p}_a,\\
 \lbrack \hat{p}_a,\hat{q}_a \rbrack &\!\!=\!\!& \rho_a,
 \hspace{20mm}
 \lbrack a_n,\hat{q}_a \rbrack = 0 ~~(n\neq 0),
\eeqa
where $\rho_a$ is a constant and $\rho_a(n)$ satisfies
\beq
 \lim_{q\rightarrow 1}\rho_a(n)=\rho_a,~~~
 \lim_{n\rightarrow 0}\rho_a(n)=
 \biggl(\frac{2\log q}{q-q^{-1}}\biggr)^2\rho_a,
\eeq
we define free boson fields $a(z;\alpha)$ and $a_{\pm}(z)$
as follows:
\beqa
 a(z;\alpha) &\!\!\eqdef\!\!& -\sum_{n\neq 0}\frac{a_n}{[n]}
 q^{-\alpha |n|}z^{-n}+\hat{q}_a+\hat{p}_a\log z, \\
 a_{\pm}(z) &\!\!\eqdef\!\!&
 \pm \Bigl( (q-q^{-1})\sum_{\pm n>0}a_n z^{-n}+\hat{p}_a\log q \Bigr) \\
 &\!\!=\!\!& \pm (q-q^{-1})
 \Bigl( \sum_{\pm n>0}a_n z^{-n}+\sfrac{1}{2}a_0 \Bigr).
\eeqa
We abbreviate $a(z;0)$ as $a(z)\eqdef a(z;0)$.
In the $q\!\rightarrow\! 1$ limit $a(z;\alpha)$ becomes the free chiral
boson field $\phi(z)$ used in the string theory and CFT
(but the meaning of $z$ is different).
Correspondence between $a(z;\alpha)$ and
$\phi(z)=\hat{x}-\sqrt{-1}\hat{p}\log z
 +\sqrt{-1}\sum_{n\neq 0}\frac{1}{n}\alpha_n z^{-n}$
is
\beq
 a(z;\alpha) \!\rightarrow\! \sqrt{-1}\sqrt{\rho_a}\phi(z),~~
 a_n \!\rightarrow\! \sqrt{\rho_a}\alpha_n,~~
 \hat{p}_a \!\rightarrow\! \sqrt{\rho_a}\hat{p},~~
 \hat{q}_a \!\rightarrow\! \sqrt{-1}\sqrt{\rho_a}\hat{x}.
\eeq

Moreover let us define boson fields with parameters $L,M$ as follows:
\beqa
 && a(L_1,\cdots,L_r;M_1,\cdots,M_r|z;\alpha) \n
 && \hspace{10mm} \eqdef
 -\sum_{n\neq 0}\frac{[L_1n]\cdots [L_rn]}{[M_1n]\cdots [M_rn]}
 \frac{a_n}{[n]}q^{-\alpha |n|}z^{-n}
 +\frac{L_1\cdots L_r}{M_1\cdots M_r}(\hat{q}_a+\hat{p}_a\log z), \\
 && a_{\pm}(L_1,\cdots,L_r;M_1,\cdots,M_r|z) \n
 && \hspace{10mm} \eqdef
 \pm \Bigl( (q-q^{-1})
 \sum_{\pm n>0}\frac{[L_1n]\cdots [L_rn]}{[M_1n]\cdots [M_rn]}a_n z^{-n}
 +\frac{L_1\cdots L_r}{M_1\cdots M_r} \hat{p}_a\log q \Bigr) \n
 && \hspace{10mm} =
 \pm (q-q^{-1}) \Bigl(
 \sum_{\pm n>0}\frac{[L_1n]\cdots [L_rn]}{[M_1n]\cdots [M_rn]}a_n z^{-n}
 +\frac{L_1\cdots L_r}{M_1\cdots M_r} \sfrac{1}{2}a_0 \Bigr).
\eeqa
We abbreviate these as
\beqa
 \Bigl( \frac{L_1}{M_1}\frac{L_2}{M_2}\cdots\frac{L_r}{M_r}
  a \Bigr)(z;\alpha)
 &\!\!\eqdef\!\!& a(L_1,L_2,\cdots,L_r;M_1,M_2,\cdots,M_r|z;\alpha), \\
 \Bigl( \frac{L_1}{M_1}\frac{L_2}{M_2}\cdots\frac{L_r}{M_r}
  a_{\pm} \Bigr)(z)
 &\!\!\eqdef\!\!& a_{\pm}(L_1,L_2,\cdots,L_r;M_1,M_2,\cdots,M_r|z).
\eeqa

Normal ordering prescription $:~~:$ is defined by
\beq
 \left\{
 \begin{array}{rcl}
  \mbox{move} & a_n~(n>0) \mbox{ and } \hat{p}_a & \mbox{ to right}, \\
  \mbox{move} & a_n~(n<0) \mbox{ and } \hat{q}_a & \mbox{ to left}.
 \end{array}
 \right.
\eeq
For example,
\beq
 :\exp\Bigl(a(z;\alpha)\Bigr):=
 \exp\Bigl( -\sum_{n<0}\frac{a_n}{[n]}(q^{-\alpha}z)^{-n} \Bigr)
 e^{\hat{q}_a}z^{\hat{p}_a}
 \exp\Bigl( -\sum_{n>0}\frac{a_n}{[n]}(q^{\alpha}z)^{-n} \Bigr).
\eeq

For multicomponent $a^i$ $(a^i_n,\hat{p}_a^i,\hat{q}_a^i)$, we treat
them similarly;
$\lbrack a^i_n,a^j_m\rbrack=n\rho_a^{ij}(n)\delta_{n+m,0}$, etc.
We can easily verify the following:
\beq
 \Bigl\lbrack \sfrac{1}{2}\sum_{i,j}\sum_{n\in \sbm{Z}}
 :a^i_{-n}\rho_a^{-1,ij}(n)a^j_n:,a^{\ell}_m \Bigr\rbrack
 =-ma^{\ell}_m,
 \Label{L02}
\eeq
where $\rho_a^{-1,ij}(n)$ is an inverse of $\rho_a^{ij}(n)$,
i.e., $\sum_{\ell}\rho_a^{i\ell}(n)\rho_a^{-1,\ell j}(n)=\delta^{ij}$.

%
%
%
%
%
\section{Free Boson Realization of $U_q(\widehat{sl_N})$}
\setcounter{equation}{0}
To construct the Drinfeld $U_q(\widehat{sl_N})$ generators of
level $k$ in terms of free boson fields,
we need $N^2-1$ free boson fields $a^i$ $(1\leq i\leq N-1)$,
$b^{ij}$ and $c^{ij}$ $(1\leq i<j\leq N)$.
Their commutation relations are
\beqa
 \lbrack a^i_n,a^j_m \rbrack
 &\!\!=\!\!&
 \frac{1}{n}[(k+g)n][a_{ij}n]\delta_{n+m,0},~~
 \lbrack \hat{p}^i_a,\hat{q}^j_a \rbrack = (k+g)a_{ij}, \\
 \lbrack b^{ij}_n,b^{i'j'}_m \rbrack
 &\!\!=\!\!&
 -\frac{1}{n}[n]^2\delta^{ii'}\delta^{jj'}\delta_{n+m,0},~~
 \hspace{6mm}
 \lbrack \hat{p}^{ij}_b,\hat{q}^{i'j'}_b \rbrack =
 -\delta^{ii'}\delta^{jj'}, \\
 \lbrack c^{ij}_n,c^{i'j'}_m \rbrack
 &\!\!=\!\!&
 \frac{1}{n}[n]^2\delta^{ii'}\delta^{jj'}\delta_{n+m,0},~~
 \hspace{9mm}
 \lbrack \hat{p}^{ij}_c,\hat{q}^{i'j'}_c \rbrack =
 \delta^{ii'}\delta^{jj'},
\eeqa
and the remaining commutators vanish.

Let us define fields $H^i(z)$, $\psi^i_{\pm}(z)$ and $E^{\pm,i}(z)$
$(1\leq i \leq N-1)$ as follows\footnote{
These operators are well-defined on the boson Fock space that
will be defined in the next section.}:
\beqa
 H^i(z) &\!\!\eqdef\!\!&
 \frac{1}{(q-q^{-1})z} \n
 && \times \Bigl(
 \sum_{j=1}^i ( b^{j,i+1}_+(q^{\frac{k}{2}+j-1}z)
               -b^{j,i}_+(q^{\frac{k}{2}+j}z)) \n
 && \hspace{5mm}
 +a^i_+(q^{\frac{g}{2}}z)
 +\sum_{j=i+1}^N ( b^{i,j}_+(q^{\frac{k}{2}+j}z)
                  -b^{i+1,j}_+(q^{\frac{k}{2}+j-1}z))
 \Bigr) \n
 &&
 -(\mbox{replacement}:x_+(q^{\alpha}z) \mapsto x_-(q^{-\alpha}z)
   \mbox{ for } x=a,b),
 \Label{Hz}\\
%
 \psi^i_{\pm}(q^{\pm\frac{k}{2}}z) &\!\!\eqdef\!\!&
 :\exp \Bigl(
 \sum_{j=1}^i ( b^{j,i+1}_{\pm}(q^{\pm(k+j-1)}z)
               -b^{j,i}_{\pm}(q^{\pm(k+j)}z)) \n
 && \hspace{10mm}
 +a^i_{\pm}(q^{\pm\frac{k+g}{2}}z)
 +\sum_{j=i+1}^N ( b^{i,j}_{\pm}(q^{\pm(k+j)}z)
                  -b^{i+1,j}_{\pm}(q^{\pm(k+j-1)}z))
 \Bigr):,
\Label{psiz}\\
%
 E^{+,i}(z) &\!\!\eqdef\!\!&
 \frac{-1}{(q-q^{-1})z} \n
 && \times
 \sum_{j=1}^i
 :\exp \Bigl( (b+c)^{j,i}(q^{j-1}z) \Bigr) \n
 && \hspace{10mm} \times \Bigl(
 \hspace{3mm} \exp (b^{j,i+1}_+(q^{j-1}z)-(b+c)^{j,i+1}(q^jz)) \n
 && \hspace{15mm}
 -\exp (b^{j,i+1}_-(q^{j-1}z)-(b+c)^{j,i+1}(q^{j-2}z))
 \Bigr) \n
 && \hspace{10mm} \times
 \exp \Bigl(
 \sum_{\ell=1}^{j-1}
 (b^{\ell,i+1}_+(q^{\ell-1}z)-b^{\ell,i}_+(q^{\ell}z))
 \Bigr):,
\Label{E+z}\\
%
 E^{-,i}(z) &\!\!\eqdef\!\!& \frac{-1}{(q-q^{-1})z} \n
 && \times \biggl(
 \sum_{j=1}^{i-1} :\exp \Bigl((b+c)^{j,i+1}(q^{-(k+j)}z) \Bigl) \n
 && \hspace{10mm} \times \Bigl(
 \hspace{3mm}
 \exp (-b^{j,i}_-(q^{-(k+j)}z)-(b+c)^{j,i}(q^{-(k+j-1)}z)) \n
 && \hspace{15mm}
 -\exp (-b^{j,i}_+(q^{-(k+j)}z)-(b+c)^{j,i}(q^{-(k+j+1)}z))
 \Bigr) \n
 && \hspace{10mm} \times \exp \Bigl(
 \sum_{\ell=j+1}^i ( b^{\ell,i+1}_-(q^{-(k+\ell-1)}z)
                    -b^{\ell,i}_-(q^{-(k+\ell)}z)) \n
 && \hspace{20mm}
 +a^i_-(q^{-\frac{k+g}{2}}z)
 +\sum_{\ell=i+1}^N ( b^{i,\ell}_-(q^{-(k+\ell)}z)
                     -b^{i+1,\ell}_-(q^{-(k+\ell-1)}z))
 \Bigr): \n
 && \hspace{4mm}
 +:\exp \Bigl( (b+c)^{i,i+1}(q^{-(k+i)}z) \Bigr) \n
 && \hspace{7mm} \times
 \exp \Bigl(
 a^i_-(q^{-\frac{k+g}{2}}z)
 +\sum_{\ell=i+1}^N
  (b^{i,\ell}_-(q^{-(k+\ell)}z)-b^{i+1,\ell}_-(q^{-(k+\ell-1)}z))
 \Bigr): \n
 && \hspace{4mm}
 -:\exp \Bigl( (b+c)^{i,i+1}(q^{k+i}z) \Bigr) \n
 && \hspace{7mm} \times
 \exp \Bigl(
 a^i_+(q^{\frac{k+g}{2}}z)
 +\sum_{\ell=i+1}^N
  (b^{i,\ell}_+(q^{k+\ell}z)-b^{i+1,\ell}_+(q^{k+\ell-1}z))
 \Bigr): \n
 && \hspace{4mm}
 -\sum_{j=i+2}^N :\exp \Bigl( (b+c)^{i,j}(q^{k+j-1}z) \Bigr) \n
 && \hspace{14mm} \times \Bigl(
 \hspace{3mm}
 \exp (b^{i+1,j}_+(q^{k+j-1}z)-(b+c)^{i+1,j}(q^{k+j}z)) \n
 && \hspace{19mm}
 -\exp (b^{i+1,j}_-(q^{k+j-1}z)-(b+c)^{i+1,j}(q^{k+j-2}z))
 \Bigr) \n
 && \hspace{14mm} \times \exp \Bigl(
 a^i_+(q^{\frac{k+g}{2}}z)
 +\sum_{\ell=j}^N ( b^{i,\ell}_+(q^{k+\ell}z)
                   -b^{i+1,\ell}_+(q^{k+\ell-1}z))
 \Bigr):
 \biggr),
 \Label{E-z}
\eeqa
where $b^{ii}\eqdef 0$, $c^{ii}\eqdef 0$ and
$(b+c)^{ij}\eqdef b^{ij}+c^{ij}$.
These expressions are guessed from free boson realizations of
$U_q(\widehat{sl_2})$\cite{S}, $U_q(\widehat{sl_3})$\cite{AOS}
and the Heisenberg realization of $U_q(sl_N)$\cite{ANO} (Appendix A).
$q$-difference expressions of these fields are given in Appendix B.
In the $q\!\rightarrow\! 1$ limit, \eqs{Hz},\eqn{E+z} and \eqn{E-z}
become the bosonized version of the Wakimoto realization of
$\widehat{sl_N}$ with level $k$\cite{W,FF2,IKKS}.

{}From \eqs{Hz} and \eqn{K}, $H^i_n$ and $K_i$ are
\beqa
 H^i_n &\!\!=\!\!&
 \sum_{j=1}^i ( b^{j,i+1}_n q^{-(\frac{k}{2}+j-1)|n|}
               -b^{j,i}_n q^{-(\frac{k}{2}+j)|n|}) \n
 &&
 +a^i_n q^{-\frac{g}{2}|n|}
 +\sum_{j=i+1}^N ( b^{i,j}_n q^{-(\frac{k}{2}+j)|n|}
                  -b^{i+1,j}_n q^{-(\frac{k}{2}+j-1)|n|}),
 \Label{Hn} \\
 K_i &\!\!=\!\!&
 q^{\sum_{j=1}^i(\hat{p}^{j,i+1}_b-\hat{p}^{j,i}_b)+\hat{p}^i_a
    +\sum_{j=i+1}^N(\hat{p}^{i,j}_b-\hat{p}^{i+1,j}_b)}.
 \Label{Ki}
\eeqa
We obtain the following proposition:
\begin{prop}
 $H^i,\psi^i_{\pm},E^{\pm,i}$ in \eqs{Hz}-\eqn{E-z} satisfy the
 relations \eqs{gamma}-\eqn{HnEm} with $\gamma=q^k$, \eq{EEE},
 and the following relations:
\beqa
 &&
 E^{+,i}(z)E^{-,j}(w)\simeq E^{-,j}(w)E^{+,i}(z) \n
 &&
 \hspace{5mm}\sim \mbox{reg.}+
 \frac{\delta^{ij}}{(q-q^{-1})w}
 \biggl(
 \frac{1}{z-q^k w}\psi^i_+(q^{\frac{k}{2}}w)
 -\frac{1}{z-q^{-k} w}\psi^i_-(q^{-\frac{k}{2}}w)
 \biggr),
 \Label{E+E-2} \\
 &&
 (z-q^{\pm a_{ij}}w)E^{\pm,i}(z)E^{\pm,j}(w)\simeq
 (q^{\pm a_{ij}}z-w)E^{\pm,j}(w)E^{\pm,i}(z)\sim\mbox{reg.},
 \Label{EzEw2} \\
 &&
 E^{\pm,i}(z)E^{\pm,j}(w)\simeq
 E^{\pm,j}(w)E^{\pm,i}(z)\sim\mbox{reg.} ~~~\mbox{for } a_{ij}=0,
 \Label{EE02}
\eeqa
where the symbol $\simeq$ and $\sim$ mean equality in the OPE sense
(in other words analytic continuation sense), and $\sim$ means
equality modulo regular parts.
\end{prop}
{\it Proof.}
A straightforward but tedious OPE calculation shows this proposition.
We give the useful formulas in Appendix C and how the poles cancel
each other in Appendix D.
For \eq{EEE} some explanation is needed.
Let us denote OPE of each term of $E^{\pm,i}(z)$ as follows
(see Appendix D for notation):
\beq
 E^{\pm,i (A)}(z) E^{\pm,j (B)}(w)\simeq
 f_{\pm}^{ijAB}(z,w)
 :E^{\pm,i (A)}(z) E^{\pm,j (B)}(w):.
\eeq
For $i=j$ there are three cases:
\beqa
 f_{\pm}^{iiAB}(z,w)
 &\!\!=\!\!&
 q^{\ell}\frac{z-w}{z-q^{\pm 2}w}
 \mbox{ and }
 f_{\pm}^{iiBA}(w,z)=
 q^{\ell}\frac{w-z}{w-q^{\pm 2}z},
 \Label{f1} \\
 f_{\pm}^{iiAB}(z,w)
 &\!\!=\!\!&
 q^{\ell}\frac{q^{\pm 2}z-w}{z-q^{\pm 2}w}
 \mbox{ and }
 f_{\pm}^{iiBA}(w,z)=
 q^{\ell},
 \Label{f2} \\
 f_{\pm}^{iiAB}(z,w)
 &\!\!=\!\!&
 q^{\ell}
 \mbox{ and }
 f_{\pm}^{iiBA}(w,z)=
 q^{\ell}\frac{q^{\pm 2}w-z}{w-q^{\pm 2}z},
 \Label{f3}
\eeqa
where $\ell\in\Sbm{Z}$ depends on $i,j,A,B,\pm$.
For $a_{ij}=-1$ there are two cases\footnote{
 For $E^{-}$, there are extra poles. However, we can discard
 them because they cancel each other.}:
\beqa
 f_{\pm}^{ijAB}(z,w)
 &\!\!=\!\!&
 q^m\frac{q^{\mp 1}z-w}{z-q^{\mp 1}w}
 \mbox{ and }
 f_{\pm}^{jiBA}(w,z)=
 q^m,
 \Label{f4} \\
 f_{\pm}^{ijAB}(z,w)
 &\!\!=\!\!&
 q^m
 \mbox{ and }
 f_{\pm}^{jiBA}(w,z)=
 q^m\frac{q^{\mp 1}w-z}{w-q^{\mp 1}z},
 \Label{f5}
\eeqa
where $m\in\Sbm{Z}$ depends on $i,j,A,B,\pm$.
These OPE equations can be translated to formal power series equations:
\beq
 E^{\pm,i (A)}(z) E^{\pm,j (B)}(w) =
 g_{\pm}^{ijAB}(z,w)
 :E^{\pm,i (A)}(z) E^{\pm,j (B)}(w):.
\eeq
\Eqs{f1}-\eqn{f5} are translated to
\beqa
 g_{\pm}^{iiAB}(z,w)
 &\!\!=\!\!&
 q^{\ell}(z-w)
 \frac{1}{z}\sum_{n\geq 0}\Bigl(q^{\pm 2}\frac{w}{z}\Bigr)^n \n
 \mbox{ and } &&
 g_{\pm}^{iiBA}(w,z)=
 q^{\ell}(w-z)
 \frac{1}{w}\sum_{n\geq 0}\Bigl(q^{\pm 2}\frac{z}{w}\Bigr)^n,
 \Label{g1} \\
 g_{\pm}^{iiAB}(z,w)
 &\!\!=\!\!&
 q^{\ell}(q^{\pm 2}z-w)
 \frac{1}{z}\sum_{n\geq 0}\Bigl(q^{\pm 2}\frac{w}{z}\Bigr)^n
 \mbox{ and }
 g_{\pm}^{iiBA}(w,z)=
 q^{\ell},
 \Label{g2} \\
 g_{\pm}^{iiAB}(z,w)
 &\!\!=\!\!&
 q^{\ell}
 \mbox{ and }
 g_{\pm}^{iiBA}(w,z)=
 q^{\ell}(q^{\pm 2}w-z)
 \frac{1}{w}\sum_{n\geq 0}\Bigl(q^{\pm 2}\frac{z}{w}\Bigr)^n,
 \Label{g3} \\
 g_{\pm}^{ijAB}(z,w)
 &\!\!=\!\!&
 q^m(q^{\mp 1}z-w)
 \frac{1}{z}\sum_{n\geq 0}\Bigl(q^{\mp 1}\frac{w}{z}\Bigr)^n
 \mbox{ and }
 g_{\pm}^{jiBA}(w,z)=
 q^m,
 \Label{g4} \\
 g_{\pm}^{ijAB}(z,w)
 &\!\!=\!\!&
 q^m
 \mbox{ and }
 g_{\pm}^{jiBA}(w,z)=
 q^m(q^{\mp 1}w-z)
 \frac{1}{w}\sum_{n\geq 0}\Bigl(q^{\mp 1}\frac{z}{w}\Bigr)^n,
 \Label{g5}
\eeqa
respectively.
A product of three $E$'s can be expressed as
\beqa
 &&
 E^{\pm,i_1(A_1)}(z_1) E^{\pm,i_2(A_2)}(z_2) E^{\pm,i_3(A_3)}(z_3) \n
 &\!\!=\!\!&
 g_{\pm}^{i_1i_2A_1A_2}(z_1,z_2)
 g_{\pm}^{i_1i_3A_1A_3}(z_1,z_3)
 g_{\pm}^{i_2i_3A_2A_3}(z_2,z_3) \n
 && \hspace{15mm} \times
 :E^{\pm,i_1(A_1)}(z_1) E^{\pm,i_2(A_2)}(z_2) E^{\pm,i_3(A_3)}(z_3):.
 \Label{EEEg}
\eeqa
We remark that this is a consequence of the bosonic realization.
Using this fact, we obtain
\beqa
 &&
 E^{\pm,i(A_1)}(z_1) E^{\pm,i(A_2)}(z_2) E^{\pm,j(B)}(w)
 -(q+q^{-1}) E^{\pm,i(A_1)}(z_1) E^{\pm,j(B)}(w) E^{\pm,i(A_2)}(z_2) \n
 &&
 +E^{\pm,j(B)}(w) E^{\pm,i(A_1)}(z_1) E^{\pm,i(A_2)}(z_2) \n
 &=\!\!&
 g_{\pm}^{iiA_1A_2}(z_1,z_2)
 \Bigl(g_{\pm}^{ijA_1B}(z_1,w)g_{\pm}^{ijA_2B}(z_2,w)
  -(q+q^{-1})g_{\pm}^{ijA_1B}(z_1,w)g_{\pm}^{jiBA_2}(w,z_2) \n
 && \hspace{10mm}
  +g_{\pm}^{jiBA_1}(w,z_1)g_{\pm}^{jiBA_2}(w,z_2)\Bigr)~\times
 :E^{\pm,i(A_1)}(z_1) E^{\pm,i(A_2)}(z_2) E^{\pm,j(B)}(w):.
\eeqa
In each case, this coefficient is antisymmetric with respect to
$z_1$ and $z_2$.
Therefore \eq{EEE} holds.
\hfill $\Box$

We remark that \eqs{E+E-2}, \eqn{EzEw2}, \eqn{EE02} imply
\eqs{E+E-}, \eqn{EzEw}, \eqn{EE0} respectively.
Therefore we obtain our main statement:
\begin{cor}
 $H^i,\psi^i_{\pm},E^{\pm,i}$ in \eqs{Hz}-\eqn{E-z} realize the
 quantum affine algebra $U_q(\widehat{sl_N})$ in the Drinfeld
 realization with $\gamma=q^k$.
\end{cor}

%
%
%
%
%
\section{Screening Currents and Vertex Operators}
\setcounter{equation}{0}
To calculate correlation functions and investigate the irreducible
representation, we need screening operators, which commute
with $U_q(\widehat{sl_N})$.
Let us define the screening currents $S^i(z)$ $(i=1,\cdots,N-1)$
as follows:
\beqa
 S^i(z) &\!\!\eqdef\!\!&
 :\exp \Bigl(
 -\Bigl(\frac{1}{k+g} a^i\Bigr)(z;\sfrac{k+g}{2})
 \Bigr):
 \tilde{S}^i(z),
 \Label{Sz} \\
 \tilde{S}^i(z)
 &\!\!\eqdef\!\!&
 \frac{-1}{(q-q^{-1})z} \n
 && \times
 \sum_{j=i+1}^N
 :\exp \Bigl( (b+c)^{i+1,j}(q^{N-j}z) \Bigr) \n
 && \hspace{12mm} \times \Bigl(
 \hspace{3mm} \exp (-b^{i,j}_-(q^{N-j}z)-(b+c)^{i,j}(q^{N-j+1}z)) \n
 && \hspace{17mm}
 -\exp (-b^{i,j}_+(q^{N-j}z)-(b+c)^{i,j}(q^{N-j-1}z))
 \Bigr) \n
 && \hspace{12mm} \times
 \exp \Bigl(
 \sum_{\ell=j+1}^N
 (b^{i+1,\ell}_-(q^{N-\ell+1}z)-b^{i,\ell}_-(q^{N-\ell}z))
 \Bigr):.
 \Label{Stz}
\eeqa
We remark that $\tilde{S}^i(z)$ is nothing else but $E^{+,N-i}(z)$
with replacement
$b^{i,j}_{\pm}\!\mapsto\! -b^{N+1-j,N+1-i}_{\mp}$,
$(b+c)^{i,j} \mapsto (b+c)^{N+1-j,N+1-i}$.
These screening currents have the following properties.
\begin{prop}
 $S^i,\tilde{S}^i$ in \eqs{Sz},\eqn{Stz} and $H^i,E^{\pm,i}$ in
 \eqs{Hz}-\eqn{E-z} satisfy the following relations:
\beqa
 \lbrack H^i_n,S^j(z)\rbrack &\!\!=\!\!& 0,
 \Label{HnSz} \\
 E^{+,i}(z)S^j(w) &\!\!\simeq\!\!&
 S^j(w)E^{+,i}(z)\sim \mbox{reg.},
 \Label{E+S} \\
 E^{-,i}(z)S^j(w) &\!\!\simeq\!\!& S^j(w)E^{-,i}(z) \n
 &\!\!\sim\!\!& \mbox{reg.}+
 \delta^{ij}{}_{k+g}\partial_w\biggl(
 \frac{1}{z-w}:\exp\Bigl(
 -\Bigl(\frac{1}{k+g}a^j\Bigr)(w;-\sfrac{k+g}{2})\Bigr):
 \biggr),
 \Label{E-S}
\eeqa
and
\beqa
 &&
 (z-q^{-a_{ij}}w)\tilde{S}^i(z)\tilde{S}^j(w)\simeq
 (q^{-a_{ij}}z-w)\tilde{S}^j(w)\tilde{S}^i(z)\sim\mbox{reg.}, \\
 &&
 \tilde{S}^i(z)\tilde{S}^j(w)\simeq
 \tilde{S}^j(w)\tilde{S}^i(z)\sim\mbox{reg.}~~~
 \mbox{for } a_{ij}=0.
\eeqa
\end{prop}
{\it Proof.}
Straightforward (see Appendices C,D).
\hfill $\Box$

\Eqs{HnSz}-\eqn{E-S} can be expressed in the commutator form.
\begin{cor}
\beqa
 \lbrack H^i_n,S^j(z)\rbrack &\!\!=\!\!& 0, \\
 \lbrack E^{+,i}_n,S^j(z)\rbrack &\!\!=\!\!& 0, \\
 \lbrack E^{-,i}_n,S^j(z)\rbrack &\!\!=\!\!&
 \delta^{ij}{}_{k+g}\partial_z\biggl(
 z^n:\exp\Bigl(
 -\Bigl(\frac{1}{k+g}a^j\Bigr)(z;-\sfrac{k+g}{2})\Bigr):
 \biggr).
\eeqa
\end{cor}
{}From this we get the desired property of the screening charges.
\begin{cor}
 If the Jackson integrals of the screening currents \eq{Sz},
\beq
 \int_0^{s\infty}S^i(z)d_pz,~~~p=q^{2(k+g)},
\eeq
are convergent, they commute with $U_q(\widehat{sl_N})$ generated
by \eqs{Hz}-\eqn{E-z}.
\end{cor}

Next we will construct the vertex operators(primary fields),
which create the $U_q(\widehat{sl_N})$ highest weight states from
the vacuum state of the boson Fock space.
The vacuum state of the boson Fock space, $|{\bf 0}\rangle$, is
defined by
\beq
 a^i_n|{\bf 0}\rangle=b^{ij}_n|{\bf 0}\rangle=
 c^{ij}_n|{\bf 0}\rangle=0~~~(n\geq 0).
\eeq
Let $|p_a,p_b,p_c\rangle$ be
\beq
 |p_a,p_b,p_c\rangle\eqdef
 \exp\Bigl(\sum_{i,j=1}^{N-1}p_a^i\frac{a^{-1}_{ij}}{k+g}\hat{q}_a^j
 +\sum_{1\leq i<j\leq N}p_b^{ij}(-1)\hat{q}_b^{ij}
 +\sum_{1\leq i<j\leq N}p_c^{ij}\hat{q}_c^{ij}\Bigr)
 |{\bf 0}\rangle,
\eeq
then $|p_a,p_b,p_c\rangle$ is the highest weight state of the boson
Fock space, i.e.,
\beqa
 &&
 a^i_n|p_a,p_b,p_c\rangle=
 b^{ij}_n|p_a,p_b,p_c\rangle=
 c^{ij}_n|p_a,p_b,p_c\rangle=0~~~(n>0), \\
 &&
 \hat{p}_a^i|p_a,p_b,p_c\rangle=p_a^i|p_a,p_b,p_c\rangle,~~~
 \hat{p}_x^{ij}|p_a,p_b,p_c\rangle=p_x^{ij}|p_a,p_b,p_c\rangle~~~
 (x=b,c).
\eeqa
The boson Fock space $F(p_a,p_b,p_c)$ is generated by
oscillators of negative mode on the highest weight state
$|p_a,p_b,p_c\rangle$. $E^{\pm,i}_n$ change $p_b-p_c$ only,
$S^i_n$ changes $p_a$ and $p_b-p_c$,
$H^i_n$ does not change $p_a$, $p_b$, $p_c$.
$|p_a,0,0\rangle$ has the following property:
\begin{prop}
 $H^i,E^{\pm,i}$ in \eqs{Hz}-\eqn{E-z} act on $|p_a,0,0\rangle$ as
 follows:
\beqa
 X_n|p_a,0,0\rangle&\!\!=\!\!&0~~~(n>0;X=H^i,E^{\pm,i}), \\
 E^{+,i}_0|p_a,0,0\rangle&\!\!=\!\!&0, \\
 H^i_0|p_a,0,0\rangle&\!\!=\!\!&p_a^i|p_a,0,0\rangle.
\eeqa
\end{prop}
{\it Proof.}
Straightforward.
$X_n$ $(n>0)$ annihilate $|p_a,p_b,p_c\rangle$ with $p_b+p_c=0$,
and $E^{+,i}_0$ annihilate $|p_a,0,0\rangle$.
\hfill $\Box$\\
This property is just the highest weight state condition of
$U_q(\widehat{sl_N})$.
\begin{cor}
 Using the highest weight state $|p_a,0,0\rangle=
 |(\ell^1,\cdots,\ell^{N-1}),0,0\rangle$, we get the highest weight
 left module of $U_q(\widehat{sl_N})$, $V(\lambda)$,
\beq
 V(\lambda)\hookrightarrow\bigoplus_{r\in\sbm{Z}^{N(N-1)/2}}
 F((\ell^1,\cdots,\ell^{N-1}),r,-r),
\eeq
 where the classical part of the highest weight is
 $\bar{\lambda}=\ell^1\bar{\Lambda}_1+\cdots
 +\ell^{N-1}\bar{\Lambda}_{N-1}=(\ell^1,\cdots,\ell^{N-1})$.
\Label{rep}
\end{cor}
As is well known in CFT, this module is reducible.

Let us define the vertex operator with a weight
$\bar{\lambda}=(\ell^1,\cdots,\ell^{N-1})$ and a parameter $\alpha$,
$\phi^{\bar{\lambda}}(z;\alpha)$, as follows:
\beq
 \phi^{\bar{\lambda}}(z;\alpha)
 \eqdef
 :\exp \Bigl(
 \sum_{i,j=1}^{N-1}
 \Bigl(
 \frac{\ell^i}{k+g} \frac{\min (i,j)}{N} \frac{N-\max (i,j)}{1} a^j
 \Bigr)(z;\alpha)
 \Bigr):.
 \Label{phi}
\eeq
The highest weight state of $U_q(\widehat{sl_N})$,
$|(\ell^1,\cdots,\ell^{N-1}),0,0\rangle$,
is created from the vacuum $|{\bf 0}\rangle$ by this operator
with any parameters $\alpha$ and $\beta$,
\beq
 |(\ell^1,\cdots,\ell^{N-1}),0,0\rangle=
 \lim_{z\rightarrow 0}
 \phi^{\bar{\lambda}}(q^\beta z;\alpha)
 |{\bf 0}\rangle.
\eeq
Moreover this vertex operator has the following properties.
\begin{prop}
 $\phi^{\bar{\lambda}}$ in \eq{phi} and $H^i,E^{\pm,i}$ in
 \eqs{Hz}-\eqn{E-z} satisfy the following relations:
\beqa
 \lbrack H^i_n, \phi^{\bar{\lambda}}(z;\alpha) \rbrack
 &\!\!=\!\!&
 \frac{1}{n}[\ell^in]q^{-(\alpha+\frac{g}{2})|n|}z^n
 \phi^{\bar{\lambda}}(z;\alpha), \\
 \lbrack E^{+,i}_n, \phi^{\bar{\lambda}}(z;\alpha)
 \rbrack
 &\!\!=\!\!& 0,
\eeqa
and
\beq
 (z-q^{\ell^i}w)E^{-,i}(z)\phi^{\bar{\lambda}}(w;-\sfrac{k+g}{2})
 \simeq(q^{\ell^i}z-w)\phi^{\bar{\lambda}}(w;-\sfrac{k+g}{2})E^{-,i}(z)
 \sim\mbox{reg.}
 \Label{E-phi}
\eeq
\end{prop}
{\it Proof.}
Straightforward.
We use the $q$-analogue of the inverse of the Cartan matrix:
\beq
 \sum_{r=1}^{N-1}
 \frac{[a_{ir}n]}{[n]}
 \frac{[\min(r,j)n][(N-\max(r,j))n]}{[Nn][n]}
 = \delta_{ij}.
 \Label{aiq}
\eeq
\hfill $\Box$\\
We remark that \eq{E-phi} can be rewritten as
\beq
 \Bigl\lbrack E^{-,i}_n, \phi^{\bar{\lambda}}(z;-\sfrac{k+g}{2})
 \Bigr\rbrack_{q^{\ell^i}}
 =-z
 \Bigl\lbrack \phi^{\bar{\lambda}}(z;-\sfrac{k+g}{2}),
 E^{-,i}_{n-1} \Bigr\rbrack_{q^{\ell^i}}.
\eeq

{}From $\phi^{\bar{\lambda}}(q^\beta z;\alpha)$ with appropriate
$\alpha$ and $\beta$, we can construct the $q$-vertex
operator $\Phi(z)$\cite{FR}, which has an intertwining property.
We will discuss this problem in the next section.

%
%
%
%
%
\section{Discussion}
\setcounter{equation}{0}
In this paper we have constructed a free boson realization of
$U_q(\widehat{sl_N})$.
We can also bosonize the grading operator $d$. $d$ is defined by
the property for the Chevalley generators,
\beq
 \lbrack d,t_i\rbrack=0,~~~
 \lbrack d,e^{\pm}_i\rbrack=\pm\delta_{i0}e^{\pm}_i,
\eeq
or equivalently, for the Drinfeld generators,
\beq
 \lbrack d,H^i_n\rbrack=nH^i_n,~~~
 \lbrack d,E^{\pm,i}_n\rbrack=nE^{\pm,i}_n.
 \Label{d}
\eeq
Using \eqs{L02} and \eqn{aiq}, let us define the $q$-analogue of
the Virasoro $L_0$ operator\cite{FF2,IKKS} as follows:
\beqa
 L_0&\!\!\eqdef\!\!&
 \sfrac{1}{2}\sum_{i,j=1}^{N-1}\sum_{n\in \sbm{Z}}
 :a^i_{-n}\frac{n^2}{[n][(k+g)n]}
 \frac{[\min(i,j)n][(N-\max(i,j))n]}{[Nn][n]}a^j_n:
 +\sum_{i,j=1}^{N-1}\bar{\rho}^i\frac{a^{-1}_{ij}}{k+g}\hat{p}_a^j \n
 &&
 +\sfrac{1}{2}\sum_{1\leq i<j\leq N}\sum_{n\in \sbm{Z}}
 :b^{ij}_{-n}(-1)\frac{n^2}{[n]^2}b^{ij}_n:
 +\sfrac{1}{2}\sum_{1\leq i<j\leq N}\hat{p}_b^{ij} \n
 &&
 +\sfrac{1}{2}\sum_{1\leq i<j\leq N}\sum_{n\in \sbm{Z}}
 :c^{ij}_{-n}\frac{n^2}{[n]^2}c^{ij}_n:
 +\sfrac{1}{2}\sum_{1\leq i<j\leq N}\hat{p}_c^{ij},
\eeqa
where $\bar{\rho}^i=1$, i.e.,
$\bar{\rho}=(1,1,\cdots,1)=\sum_{i=1}^{N-1}\bar{\Lambda}_i$ is the
half sum of positive roots of $sl_N$. Then $d=-L_0$ satisfies \eq{d}
on the representation space given in Cor.\ref{rep}.
The $L_0$ eigenvalue of $|(\ell^1,\cdots,\ell^{N-1}),0,0\rangle$ is
$\frac{1}{2(k+g)}\ell^ia^{-1}_{ij}(\ell^j+2\bar{\rho}^j)$$=$
$\frac{1}{2(k+g)}(\bar{\lambda},\bar{\lambda}+2\bar{\rho})$.

We have also constructed the screening currents and the vertex
operators. Using these, we can start the representation theory and
calculation of correlation functions. Like as $\widehat{sl_N}$
\cite{BF,BMP}, it is expected that the projection from the boson Fock
space to the irreducible $U_q(\widehat{sl_N})$ representation space
can be done by BRST cohomology technique.
In fact, recently, this procedure has been worked out for
$U_q(\widehat{sl_2})$\cite{K}.
The BRST operator is constructed by using the screening current.

To calculate the Jackson integral formulas for the correlation
functions, which are solutions of $q$-KZ equation,
we must first prepare the $q$-vertex operators $\Phi$.
We will restrict ourselves to the
type \uppercase\expandafter{\romannumeral 1} \cite{DFJMN} vertex
operator $\Phi_{V(\mu)}^{V(\nu)V_{\lambda}}(z):V(\mu)\rightarrow
V(\nu)\otimes V_{\lambda z}$.
$\Phi_{V(\mu)}^{V(\nu)V_{\lambda}}(z)$ can be constructed
from $\phi^{\bar{\lambda}}(q^{\beta}z;\alpha)$ with appropriate
$\alpha$, $\beta$. From \eq{E-phi}, we choose $\alpha=-\frac{k+g}{2}$.
This choice agrees with refs.\cite{KQS} ($U_q(\widehat{sl_2})$ with
an arbitrary level $k$) and \cite{Ko} (vector representation of
$U_q(\widehat{sl_N})$ with $k=1$).
Starting from $\phi^{\bar{\lambda}}(z)\eqdef
\phi^{\bar{\lambda}}(z;-\frac{k+g}{2})$, we define
$\phi^{\bar{\lambda}}_{i_1,\cdots,i_n}(z)$ as follows:
\beq
 \phi^{\bar{\lambda}}_{i_1,\cdots,i_n}(z)
 \eqdef \Bigl\lbrack
 \phi^{\bar{\lambda}}_{i_1,\cdots,i_{n-1}}(z),E^{-,i_n}_0
 \Bigr\rbrack_{q^x},~~~
 x=(\bar{\lambda}-\sum_{j=1}^{n-1}\bar{\alpha}_{i_j},
    \bar{\alpha}_{i_n}).
\eeq
To determine $\beta$, we must the specify finite dimensional
representation of $U_q(\widehat{sl_N})$.
Results of refs.\cite{KQS,Ko} suggest $\beta=k+g$. Once the
finite dimensional representation is obtained and $\beta$ is
determined, we can construct the $q$-vertex operator
$\Phi_{V(\mu)}^{V(\nu)V_{\lambda}}(z)$ from our
$\phi^{\bar{\lambda}}_{i_1,\cdots,i_n}(q^{\beta}z)$.
Then, we can calculate correlation functions of the $q$-vertex
operators in standard way. These problems are now under investigation.

To extend our results to arbitrary quantum affine Lie algebras,
it may be important to consider the geometrical interpretation of
the free boson realization. For $q=1$ case, the $\beta$-$\gamma$
system is suitable for the geometrical interpretation\cite{FF2}.
For $q\neq 1$ case, we define the quantum $\beta$-$\gamma$ fields,
$\beta_{\alpha,\pm}^{ij}(z)$ ($\alpha=\pm 1$), $\gamma^{ij}(z)$,
as follows:
\beqa
 \beta_{1,\pm}(z)&\!\!\eqdef\!\!&
 \frac{-1}{(q-q^{-1})z}:\exp\Bigl(
 b_{\pm}(z)-(b+c)(q^{\pm 1}z)\Bigl):, \\
 \beta_{-1,\pm}(z)&\!\!\eqdef\!\!&
 \frac{-1}{(q-q^{-1})z}:\exp\Bigl(
 -b_{\mp}(z)-(b+c)(q^{\pm 1}z)\Bigl):, \\
 \gamma(z)&\!\!\eqdef\!\!&
 :\exp\Bigl((b+c)(z)\Bigl):,
\eeqa
where we suppress the superscript $ij$.
They are not free fields any longer. They satisfy
\beqa
 (z-q^{\alpha+\alpha'}w)\beta_{\alpha,\epsilon}(z)
 \beta_{\alpha',\epsilon'}(w)
 &\!\!=\!\!&
 (q^{\alpha+\alpha'}z-w)\beta_{\alpha',\epsilon'}(w)
 \beta_{\alpha,\epsilon}(z)~~~~~(\epsilon,\epsilon'=\pm), \\
 \beta_{\pm 1,\pm}(z)\beta_{\mp 1,\pm}(w)
 &\!\!=\!\!&
 \beta_{\mp 1,\pm}(w)\beta_{\pm 1,\pm}(z), \\
 (z-q^{\mp 1}w)\beta_{\pm 1,\pm}(z)\gamma(w)
 &\!\!=\!\!&
 (q^{\mp 1}z-w)\gamma(w)\beta_{\pm 1,\pm}(z), \\
 (z-q^{\mp 1}w)\gamma(z)\beta_{\pm1,\mp}(w)
 &\!\!=\!\!&
 (q^{\mp 1}z-w)\beta_{\pm 1,\mp}(w)\gamma(z), \\
 \gamma(z)\gamma(w)
 &\!\!=\!\!&
 \gamma(w)\gamma(z).
\eeqa
Our free boson realization of $U_q(\widehat{sl_N})$ is reexpressed
by these quantum $\beta$-$\gamma$ fields.
In the $q\rightarrow 1$ limit,
$\beta_{\alpha,+}(z)-\beta_{\alpha,-}(z)$ and $\gamma(z)$ become
usual $\beta(z)$ and $\gamma(z)$ respectively.
These $\beta_{\alpha,\pm}$, $\gamma$ fields are the affinization
of $q$-oscillator($aa^{\dagger}-q^{\pm 1}a^{\dagger}a=q^{\mp{\cal N}}$);
$a\rightarrow\gamma$,
$a^{\dagger}\rightarrow\beta_{\alpha,+}-\beta_{\alpha,-}$
(see Appendix A).
We expect that our realization in terms of the quantum $\beta$-$\gamma$
system acts on the $q$-deformed semi-infinite flag manifold\cite{FF2}.

Our free boson realization may be also useful to analyze the
$q$-analogue of the Virasoro and $W$ algebras by the Hamiltonian
reduction, and the representation at the critical level $k=-g$.

\section*{Acknowledgments}
\noindent
We would like to thank T. Inami for a careful reading
the manuscript and discussions.
We would also like to thank T. Eguchi, E. Frenkel, M. Jimbo,
T. Miwa, A. Nakayashiki, M. Noumi, and Y. Yamada
for helpful discussions.

\newpage
\section* {Appendix A}
\setcounter{section}{1}
\renewcommand{\thesection}{\Alph{section}}
\setcounter{equation}{0}
For the reader's convenience, we give the result of \cite{ANO},
the Heisenberg realization of $U_q(sl_N)$ with the weight
$\lambda_i\in\Sbm{C}$.
Let us consider variables $x_{ij}$ and
derivatives $\frac{\partial}{\partial x_{ij}}$ $(1\leq i<j\leq N)$.
Their commutation relations are
\beq
 \Bigl\lbrack
 \frac{\partial}{\partial x_{ij}},x_{i'j'}
 \Bigr \rbrack
 =\delta_{ii'}\delta_{jj'},~~~
 \mbox{others}=0.
\eeq
Standard Chevalley generators of $U_q(sl_N)$, $e^{\pm}_i$,
$t_i=q^{h_i}$ $(i=1,\cdots,N-1)$, are realized as follows:
\beqa
 h_i &\!\!\eqdef\!\!&
 -\sum_{j=1}^i(\vartheta_{j,i+1}-\vartheta_{j,i})
 +\lambda_i
 -\sum_{j=i+1}^{N}(\vartheta_{i,j}-\vartheta_{i+1,j}), \\
 e^+_i &\!\!\eqdef\!\!&
 \sum_{j=1}^ix_{j,i}\frac{1}{x_{j,i+1}}[\vartheta_{j,i+1}]
 q^{-\sum_{\ell=1}^{j-1}(\vartheta_{\ell,i+1}-\vartheta_{\ell,i})}, \\
 e^-_i &\!\!\eqdef\!\!&
 \sum_{j=1}^{i-1}x_{j,i+1}\frac{1}{x_{j,i}}[\vartheta_{j,i}]
 q^{\sum_{\ell=j+1}^i(\vartheta_{\ell,i+1}-\vartheta_{\ell,i})
    -\lambda_i
    +\sum_{\ell=i+1}^N(\vartheta_{i,\ell}-\vartheta_{i+1,\ell})} \n
 &&
 +x_{i,i+1}[\lambda_i
  -\sum_{\ell=i+1}^N(\vartheta_{i,\ell}-\vartheta_{i+1,\ell})] \n
 &&
 -\sum_{j=i+2}^Nx_{i,j}\frac{1}{x_{i+1,j}}[\vartheta_{i+1,j}]
 q^{\lambda_i
    -\sum_{\ell=j}^N(\vartheta_{i,\ell}-\vartheta_{i+1,\ell})},
\eeqa
where $\vartheta_{ij}\eqdef x_{ij}\frac{\partial}{\partial x_{ij}}$,
$x_{ii}\eqdef 1$, $\vartheta_{ii}\eqdef 0$.

Our free field realization of $U_q(\widehat{sl_N})$ is obtained by
the following replacement with suitable argument:
\beqa
 x &\mapsto& e^{(b+c)(z)},\\
 -\vartheta &\mapsto& \pm b_{\pm}(z),\\
 \lambda &\mapsto& \pm a_{\pm}(z),\\
 \lbrack A(z)\rbrack &\mapsto&
 \frac{e^{A_+(z)}-e^{A_-(z)}}{(q-q^{-1})z}.
\eeqa
%
%
%
%
%
%
\section* {Appendix B}
\setcounter{section}{2}
\setcounter{equation}{0}
In this appendix, we reexpress \eqs{Hz},\eqn{E+z},\eqn{E-z} and
\eqn{Sz} by using the $q$-difference operator.
These expressions are not unique and we give one of them.

Using the following formulas
\beqa
 &&
 \frac{1}{(q-q^{-1})z}
 \Bigl(a_+(q^{\alpha}z)-a_-(q^{-\alpha}z)\Bigr)
 = {}_1\partial_z a(z;\alpha)
 = \sum_{n\in \sbm{Z}}a_nq^{-\alpha|n|}z^{-n-1}, \\
 &&
 \frac{1}{(q-q^{-1})z}
 :\Bigl(\exp(\pm b_{\pm}(z)-(b+c)(qz))
 -\exp(\pm b_{\mp}(z)-(b+c)(q^{-1}z))\Bigr): \n
 && \hspace{20mm}
 = :{}_1\partial_z \Bigl( \exp(-c(z)) \Bigr)
 \cdot \exp(-b(z;\mp 1)):, \\
 &&
 \frac{1}{(q-q^{-1})z}
 :\Bigl(\exp(\pm a_+(q^{\alpha}z))
 -\exp(\pm a_-(q^{-\alpha}z))\Bigr): \n
 && \hspace{20mm}
 = :{}_M\partial_z \biggl(
 \exp\Bigl(\pm\Bigl(\frac{1}{M}a\Bigr)(z;\alpha)\Bigr) \biggr)
 \cdot \exp\Bigl(\mp\Bigl(\frac{1}{M}a\Bigr)(z;\alpha-M)\Bigr):, \\
 &&
 \frac{1}{(q-q^{-1})z}
 :\Bigl(\exp(b(q^{\alpha}z))
 -\exp(b(q^{-\alpha}z))\Bigr): \n
 && \hspace{20mm}
 = :{}_M\partial_z \biggl(
 \exp\Bigl(\Bigl(\frac{\alpha}{M}b\Bigr)(z)\Bigr) \biggr)
 \cdot \exp\Bigl(\Bigl(\frac{M-\alpha}{M}b\Bigr)(z)\Bigr):, \\
 &&
 \Bigl( \frac{\alpha}{M}b \Bigr)(z)\pm
 \Bigl( \frac{1}{M}b \Bigr)(z;\pm\alpha+1)
 =
 \Bigl( \frac{\alpha\pm 1}{M}b \Bigr)(z;1),
\eeqa
\eqs{Hz},\eqn{E+z},\eqn{E-z} and \eqn{Sz} are rewritten as follows:
\beqa
 H^i(z) &\!\!=\!\!&
 {}_1\partial_z \Bigl(
 \sum_{j=1}^i ( b^{j,i+1}(z;\sfrac{k}{2}+j-1)
               -b^{j,i}(z;\sfrac{k}{2}+j)) \n
 && \hspace{7mm}
 +a^i(z;\sfrac{g}{2})
 +\sum_{j=i+1}^N ( b^{i,j}(z;\sfrac{k}{2}+j)
                  -b^{i+1,j}(z;\sfrac{k}{2}+j-1))
 \Bigr), \\
 E^{+,i}(z) &\!\!=\!\!&
 -\sum_{j=1}^i
 :\exp \Bigl( (b+c)^{j,i}(q^{j-1}z) \Bigr) \n
 &&\hspace{8mm}
 \times {}_1\partial_z
 \Bigl( \exp (-c^{j,i+1}(q^{j-1}z))\Bigr) \cdot
 \exp ( -b^{j,i+1}(q^{j-1}z;-1)) \n
 && \hspace{8mm} \times
 \exp \Bigl(
 \sum_{\ell=1}^{j-1}
 (b^{\ell,i+1}_+(q^{\ell-1}z)-b^{\ell,i}_+(q^{\ell}z))
 \Bigr):, \\
 E^{-,i}(z) &\!\!=\!\!&
 -\sum_{j=1}^{i-1} :\exp \Bigl((b+c)^{j,i+1}(q^{-(k+j)}z) \Bigr) \n
 &&\hspace{8mm}
 \times {}_1\partial_z \Bigl( \exp(-c^{ji}(q^{-(k+j)}z)) \Bigr)
 \cdot \exp(-b^{ji}(q^{-(k+j)}z;1)) \n
 && \hspace{8mm} \times \exp \Bigl(
 \sum_{\ell=j+1}^i ( b^{\ell,i+1}_-(q^{-(k+\ell-1)}z)
                    -b^{\ell,i}_-(q^{-(k+\ell)}z)) \n
 && \hspace{20mm}
 +a^i_-(q^{-\frac{k+g}{2}}z)
 +\sum_{\ell=i+1}^N ( b^{i,\ell}_-(q^{-(k+\ell)}z)
                     -b^{i+1,\ell}_-(q^{-(k+\ell-1)}z))
 \Bigr): \n
 &&
 +:{}_{k+g}\partial_z\biggl(
 \exp \Bigl(
 \Bigl(\frac{k+i}{k+g}(b+c)^{i,i+1}\Bigr)(z)
 +\Bigl(\frac{1}{k+g}a^i\Bigr)(z;\sfrac{k+g}{2}) \n
 && \hspace{18mm}
 +\sum_{\ell=i+1}^N\Bigl(
  \Bigl(\frac{1}{k+g}b^{i,\ell}\Bigr)(z;k+\ell)
  -\Bigl(\frac{1}{k+g}b^{i+1,\ell}\Bigr)(z;k+\ell-1)\Bigr)
 \Bigr)\biggr) \n
 && \hspace{3mm}
 \times
 \exp \Bigl(
 \Bigl(\frac{g-i}{k+g}(b+c)^{i,i+1}\Bigr)(z)
 -\Bigl(\frac{1}{k+g}a^i\Bigr)(z;-\sfrac{k+g}{2}) \n
 && \hspace{15mm}
 -\sum_{\ell=i+1}^N\Bigl(
  \Bigl(\frac{1}{k+g}b^{i,\ell}\Bigr)(z;\ell-g)
  -\Bigl(\frac{1}{k+g}b^{i+1,\ell}\Bigr)(z;\ell-g-1)\Bigr)
 \Bigr): \n
 &&
 +\sum_{j=i+2}^N :\exp \Bigl( (b+c)^{i,j}(q^{k+j-1}z) \Bigr) \n
 &&\hspace{12mm}
 \times
 {}_1\partial_z\Bigl(\exp(-c^{i+1,j}(q^{k+j-1}z))\Bigr)
 \cdot \exp(-b^{i+1,j}(q^{k+j-1}z;-1)) \n
 && \hspace{12mm} \times \exp \Bigl(
 a^i_+(q^{\frac{k+g}{2}}z)
 +\sum_{\ell=j}^N ( b^{i,\ell}_+(q^{k+\ell}z)
                   -b^{i+1,\ell}_+(q^{k+\ell-1}z))
 \Bigr):, \\
 S^i(z) &\!\!=\!\!&
 -:\exp\Bigl(-\Bigl(\frac{1}{k+g}a^i\Bigr)(z;\sfrac{k+g}{2})\Bigr) \n
 && \times
 \sum_{j=i+1}^N
 \exp \Bigl( (b+c)^{i+1,j}(q^{N-j}z) \Bigr) \n
 &&\hspace{10mm}
 \times {}_1\partial_z
 \Bigl( \exp (-c^{i,j}(q^{N-j}z))\Bigr) \cdot
 \exp ( -b^{i,j}(q^{N-j}z;1)) \n
 && \hspace{10mm} \times
 \exp \Bigl(
 \sum_{\ell=j+1}^{N}
 (b^{i+1,\ell}_-(q^{N-\ell+1}z)-b^{i,\ell}_-(q^{N-\ell}z))
 \Bigr):.
\eeqa

These expressions are adequate for taking the $q\rightarrow 1$ limit,
because there is no denominator $q-q^{-1}$.
In this limit ${}_{\alpha}\partial_z$,
$(\frac{L_1}{M_1}\cdots\frac{L_r}{M_r}a)(z;\alpha)$,
$(\frac{L_1}{M_1}\cdots\frac{L_r}{M_r}a_{\pm})(z)$ become
$\alpha\partial_z$, $\frac{L_1\cdots L_r}{M_1\cdots M_r}a(z)$, $0$
respectively.
\Eqs{Hz},\eqn{E+z} and \eqn{E-z} become the bosonized
version of the Wakimoto realization of $\widehat{sl_N}$ with
level $k$\cite{W,FF2,IKKS};
$\beta^{ij}(z)$ and $\gamma^{ij}(z)$ are expressed in terms
of $b^{ij}(z)$ and $c^{ij}(z)$ with $q=1$ as follows\cite{FMS}:
\beqa
 \beta^{ij}(z)&\!\!=\!\!&
 -:\partial_z\biggl(\exp\Bigl(-c^{ij}(z)\Bigr)\biggr)\cdot
 \exp\Bigl(-b^{ij}(z)\Bigr):, \\
 \gamma^{ij}(z)&\!\!=\!\!&
 :\exp\Bigl((b+c)^{ij}(z)\Bigr):.
\eeqa

%
%
%
%
%
%
\section* {Appendix C}
\setcounter{section}{3}
\setcounter{equation}{0}
In this appendix we give useful formulas.

First we give formulas for a boson $a$ in section 2
(see the footnote below \eq{K}).
\beqa
 \lbrack A,B\rbrack \mbox{ commute with } A,B
 &\Rightarrow&
 \lbrack A,e^B\rbrack=\lbrack A,B\rbrack e^B, \\
 &&
 e^Ae^B=e^{\lbrack A,B\rbrack}e^Be^A,
\eeqa
\beqa
 \biggl\lbrack
 a_n,\Bigl(\frac{L_1}{M_1}\cdots\frac{L_r}{M_r}a_{\pm}\Bigr)(z)
 \biggr\rbrack
 &\!\!=\!\!&
 \pm\theta(\mp n>0)(q-q^{-1})
 \frac{[L_1n]\cdots[L_rn]}{[M_1n]\cdots[M_rn]}
 n\rho_a(n)z^n, \\
 \biggl\lbrack
 a_n,\Bigl(\frac{L_1}{M_1}\cdots\frac{L_r}{M_r}a\Bigr)(z;\alpha)
 \biggr\rbrack
 &\!\!=\!\!&
 \frac{[L_1n]\cdots[L_rn]}{[M_1n]\cdots[M_rn]}
 \frac{n}{[n]}\rho_a(n)q^{-\alpha|n|}z^n,
\eeqa
\beqa
 &&
 \biggl\lbrack
 \Bigl(\frac{L_1}{M_1}\cdots\frac{L_r}{M_r}a_+\Bigr)(z),
 \Bigl(\frac{L_1'}{M_1'}\cdots\frac{L_s'}{M_s'}a_-\Bigr)(w)
 \biggr\rbrack \n
 && \hspace{20mm}
 =
 -(q-q^{-1})^2\sum_{n>0}
 \frac{[L_1n]\cdots[L_rn]}{[M_1n]\cdots[M_rn]}
 \frac{[L_1'n]\cdots[L_s'n]}{[M_1'n]\cdots[M_s'n]}
 n\rho_a(n)\Bigl(\frac{w}{z}\Bigr)^n, \\
 &&
 \biggl\lbrack
 \Bigl(\frac{L_1}{M_1}\cdots\frac{L_r}{M_r}a\Bigr)(z;\alpha),
 \Bigl(\frac{L_1'}{M_1'}\cdots\frac{L_s'}{M_s'}a_-\Bigr)(w)
 \biggr\rbrack \n
 && \hspace{20mm}
 =
 (q-q^{-1})\sum_{n>0}
 \frac{[L_1n]\cdots[L_rn]}{[M_1n]\cdots[M_rn]}
 \frac{[L_1'n]\cdots[L_s'n]}{[M_1'n]\cdots[M_s'n]}
 \frac{n}{[n]}\rho_a(n)q^{-\alpha n}\Bigl(\frac{w}{z}\Bigr)^n \n
 && \hspace{25mm}
 +\frac{L_1\cdots L_r}{M_1\cdots M_r}
 \frac{L_1'\cdots L_s'}{M_1'\cdots M_s'}
 \rho_a\log q,
 \Label{aa-} \\
 &&
 \biggl\lbrack
 \Bigl(\frac{L_1'}{M_1'}\cdots\frac{L_s'}{M_s'}a_+\Bigr)(z),
 \Bigl(\frac{L_1}{M_1}\cdots\frac{L_r}{M_r}a\Bigr)(w;\alpha)
 \biggr\rbrack
 =
 \mbox{\eq{aa-}}, \\
 &&
 \biggl\lbrack
 \Bigl(\frac{L_1}{M_1}\cdots\frac{L_r}{M_r}a\Bigr)(z;\alpha),
 \Bigl(\frac{L_1'}{M_1'}\cdots\frac{L_s'}{M_s'}a\Bigr)(w;\beta)
 \biggr\rbrack \n
 && \hspace{20mm}
 =
 -\sum_{n\neq 0}
 \frac{[L_1n]\cdots[L_rn]}{[M_1n]\cdots[M_rn]}
 \frac{[L_1'n]\cdots[L_s'n]}{[M_1'n]\cdots[M_s'n]}
 \frac{n}{[n]^2}\rho_a(n)
 q^{-(\alpha+\beta)|n|}\Bigl(\frac{w}{z}\Bigr)^n \n
 && \hspace{25mm}
 -\frac{L_1\cdots L_r}{M_1\cdots M_r}
 \frac{L_1'\cdots L_s'}{M_1'\cdots M_s'}
 \rho_a\log\frac{w}{z},
\eeqa
where $\theta(P)$ is a step function, $\theta(P)=1(0)$ when the
proposition $P$ is true(false). These are formal power series equations.

Next we give specific formulas often used in proofs.
For calculation of $\lbrack H^i_n,*\rbrack$,
\beqa
 \lbrack a^i_n,a^j_{\pm}(z)\rbrack &\!\!=\!\!&
 \pm\theta(\mp n>0)(q-q^{-1})\frac{1}{n}[(k+g)n][a_{ij}n]z^n, \\
 \biggl\lbrack
 a^i_n,\Bigl(\frac{1}{k+g}a^j\Bigr)(z;\alpha)
 \biggr\rbrack
 &\!\!=\!\!&
 \frac{1}{n}[a_{ij}n]q^{-\alpha |n|}z^n, \\
 \lbrack b_n,b_{\pm}(z)\rbrack &\!\!=\!\!&
 \mp\theta(\mp n>0)(q-q^{-1})\frac{1}{n}[n]^2z^n, \\
 \lbrack b_n,b(z)\rbrack &\!\!=\!\!&
 -\frac{1}{n}[n]z^n,
\eeqa
where we suppress the superscript of $b^{ij}$. For OPE calculation,
\beqa
 &&
 \exp\Bigl(\alpha b_+(z)\Bigr)
 \exp\Bigl(\beta b_-(w)\Bigr) \n
 &\!\!\simeq\!\!&
 \biggl(\frac{(z-w)^2}{(z-q^2w)(z-q^{-2}w)}\biggr)^{\alpha\beta}
 \exp\Bigl(\beta b_-(w)\Bigr)
 \exp\Bigl(\alpha b_+(z)\Bigr),
\eeqa
\beqa
 \exp\Bigl(\alpha b_+(z)\Bigr)
 :\exp\Bigl(\beta b(w)\Bigr):
 &\!\!\simeq\!\!&
 \biggl(\frac{z-qw}{qz-w}\biggr)^{\alpha\beta}
 :\exp\Bigl(\beta b(w)\Bigr):
 \exp\Bigl(\alpha b_+(z)\Bigr), \\
 :\exp\Bigl(\alpha b(z)\Bigr):
 \exp\Bigl(\beta b_-(w)\Bigr)
 &\!\!\simeq\!\!&
 \biggl(\frac{z-qw}{qz-w}\biggr)^{\alpha\beta}
 \exp\Bigl(\beta b_-(w)\Bigr)
 :\exp\Bigl(\alpha b(z)\Bigr): \n
 &\!\!=\!\!&
 \biggl(\frac{z-qw}{qz-w}\biggr)^{\alpha\beta}
 q^{\alpha\beta}
 :\exp\Bigl(\alpha b(z)+\beta b_-(w)\Bigr):,
\eeqa
\beqa
 &&
 \exp\Bigl(a^i_+(q^{\frac{k+g}{2}}z)\Bigr)
 \exp\Bigl(a^j_-(q^{-\frac{k+g}{2}}w)\Bigr) \n
 &\!\!\simeq\!\!&
 \frac{z-q^{a_{ij}}w}{z-q^{-a_{ij}}w}
 \frac{z-q^{-a_{ij}-2(k+g)}w}{z-q^{a_{ij}-2(k+g)}w}
 \exp\Bigl(a^j_-(q^{-\frac{k+g}{2}}w)\Bigr)
 \exp\Bigl(a^i_+(q^{\frac{k+g}{2}}z)\Bigr), \\
 &&
 \exp\Bigl(a^i_+(q^{\frac{k+g}{2}}z)\Bigr)
 :\exp\Bigl(-\Bigl(\frac{1}{k+g}a^j\Bigr)(w;\sfrac{k+g}{2})
 \Bigr): \n
 &\!\!\simeq\!\!&
 \frac{z-q^{a_{ij}-(k+g)}w}{q^{a_{ij}}z-q^{-(k+g)}w}
 :\exp\Bigl(-\Bigl(\frac{1}{k+g}a^j\Bigr)(w;\sfrac{k+g}{2})\Bigr):
 \exp\Bigl(a^i_+(q^{\frac{k+g}{2}}z)\Bigr), \\
 &&
 :\exp\Bigl(-\Bigl(\frac{1}{k+g}a^i\Bigr)(z;\sfrac{k+g}{2})\Bigr):
 \exp\Bigl(a^j_-(q^{-\frac{k+g}{2}}w)\Bigr) \n
 &\!\!\simeq\!\!&
 \frac{z-q^{a_{ij}-(k+g)}w}{q^{a_{ij}}z-q^{-(k+g)}w}
 \exp\Bigl(a^j_-(q^{-\frac{k+g}{2}}w)\Bigr)
 :\exp\Bigl(-\Bigl(\frac{1}{k+g}a^i\Bigr)(z;\sfrac{k+g}{2})\Bigr):
 \n
 &\!\!=\!\!&
 \frac{z-q^{a_{ij}-(k+g)}w}{q^{a_{ij}}z-q^{-(k+g)}w}
 q^{a_{ij}}
 :\exp\Bigl(-\Bigl(\frac{1}{k+g}a^i\Bigr)(z;\sfrac{k+g}{2})
 +a^j_-(q^{-\frac{k+g}{2}}w)\Bigr):,
\eeqa
where $\alpha$ and $\beta$ are parameters and
$\simeq$ means equality in the OPE sense (analytic continuation
sense).\\
$\exp(b+c)$'s commute each other because $\rho_b(n)+\rho_c(n)=0$.

%
%
%
%
%
%
\section* {Appendix D}
\setcounter{section}{4}
\setcounter{equation}{0}
In this appendix we give how poles cancel each other in OPE
of $E^{+,i}(z)$ and $E^{-,j}(w)$, $E^{-,i}(z)$ and $E^{-,j}(w)$,
$E^{\pm,i}(z)$ and $S^j(w)$.
Let us denote each term of \eqs{E+z},\eqn{E-z},\eqn{Sz} as
follows\footnote{
 For example, $E^{+,i(j,2)}(z)=\frac{-1}{(q-q^{-1})z}
 :\exp \Bigl( (b+c)^{j,i}(q^{j-1}z) \Bigr)
 \times (-1)
 \exp\Bigl(b^{j,i+1}_-(q^{j-1}z)-(b+c)^{j,i+1}(q^{j-2}z)\Bigr)
 \times
 \exp \Bigl(\sum_{\ell=1}^{j-1}
 (b^{\ell,i+1}_+(q^{\ell-1}z)-b^{\ell,i}_+(q^{\ell}z))\Bigr):$.
}:
\beqa
 E^{+,i}(z) &\!\!=\!\!&
 \sum_{j=1}^i(E^{+,i(j,1)}(z)+E^{+,i(j,2)}(z)), \\
 E^{-,i}(z) &\!\!=\!\!&
 \sum_{j=1}^{i-1}(E^{-,i(j,1)}(z)+E^{-,i(j,2)}(z)) \n
 &&
 +E^{-,i(i,1)}(z)+E^{-,i(i,2)}(z)
 +\sum_{j=i+2}^N(E^{-,i(j,1)}(z)+E^{-,i(j,2)}(z)), \\
 S^i(z) &\!\!=\!\!&
 \sum_{j=i+1}^N(S^{i(j,1)}(z)+S^{i(j,2)}(z)).
\eeqa

\noindent
{\bf \uppercase\expandafter{\romannumeral 1}.}
$E^{+,i}(z)E^{-,j}(w)$. \\
For $i=j$, OPE $E^{+,i}(z)E^{-,j}(w)$ has poles at $z=q^kw$
and $z=q^{-k}w$. They come from $E^{+,i(i,1)}(z)E^{-,j(j,2)}(w)$
and $E^{+,i(1,2)}(z)E^{-,j(1,1)}(w)$ respectively.\\
Some terms of $E^{+,i}(z)E^{-,j}(w)$ have other poles but all these
poles cancel in pairs. We give these poles($z=q^{\alpha}w$) and
pairs ($E^{+,i(A)}(z)E^{-,j(B)}(w)$ and $E^{+,i(C)}(z)E^{-,j(D)}(w)$).
$$
\begin{array}{lllllll}
 &&\alpha&(A)&(B)&(C)&(D)\\
 (\mbox{\romannumeral1})&j=i&-k-2\ell&(\ell,1)&(\ell,2)&(\ell+1,2)
    &(\ell+1,1)\\
 &&&&&&1\leq\ell\leq i-1\\
 (\mbox{\romannumeral2})&j=i+1&-k-2\ell+1&(\ell,1)&(\ell,2)&(\ell,2)
    &(\ell,1)\\
 &&&&&&1\leq\ell\leq i\\
 (\mbox{\romannumeral3})&j=i-1&k+1&(i-1,1)&(j+2,1)&(i,1)&(j,2)\\
 &&k+1&(i-1,1)&(j+2,2)&(i,2)&(j,2)\\
 &&k+1&(i,1)&(j+2,2)&(i,2)&(j+2,1)\\
 (\mbox{\romannumeral4})&j\leq i-2&k+i-j&(j,1)&(i+1,1)&(j+1,1)&(i,1)\\
 &&k+i-j&(j,1)&(i+1,2)&(j+1,2)&(i,1)\\
 &&k+i-j&(j+1,1)&(i+1,2)&(j+1,2)&(i+1,1).
\end{array}
$$

\noindent
{\bf \uppercase\expandafter{\romannumeral 2}.}
$E^{-,i}(z)E^{-,j}(w)$. \\
$E^{-,i}(z)E^{-,j}(w)$ has poles at $z=q^{-a_{ij}}w$.
Some terms of this OPE have extra poles. But these extra poles
($z=q^{\alpha}w$) cancel in the following pairs
($E^{-,i(A)}(z)E^{-,j(B)}(w)$ and $E^{-,i(C)}(z)E^{-,j(D)}(w)$).
$$
\begin{array}{lllllll}
 &&\alpha&(A)&(B)&(C)&(D)\\
 (\mbox{\romannumeral1})&j=i-1&2k+i+j&(i-1,2)&(j,2)&(i,1)&(j+2,2)\\
 (\mbox{\romannumeral2})&j\leq i-2&2k+i+j&(j,2)&(i,1)&(j+1,1)&(i+1,2)\\
 &&2k+i+j&(j,2)&(i,2)&(j+1,2)&(i+1,2)\\
 &&2k+i+j&(j+1,2)&(i,1)&(j+1,1)&(i,2).
\end{array}
$$

\noindent
{\bf \uppercase\expandafter{\romannumeral 3}.}
$E^{+,i}(z)S^j(w)$. \\
Poles ($z=q^{\alpha}w$) cancel in the following pairs
($E^{+,i(A)}(z)S^{j(B)}(w)$ and $E^{+,i(C)}(z)S^{j(D)}(w)$).
$$
\begin{array}{lllllll}
 &&\alpha&(A)&(B)&(C)&(D)\\
 (\mbox{\romannumeral1})&j=i&N-i-j&(i,1)&(j+1,2)&(i,2)&(j+1,1)\\
 (\mbox{\romannumeral2})&j\leq i-1&N-i-j&(j,1)&(i,2)&(j+1,2)&(i+1,1)\\
 &&N-i-j&(j,1)&(i+1,2)&(j,2)&(i+1,1)\\
 &&N-i-j&(j,2)&(i,2)&(j+1,2)&(i+1,2).
\end{array}
$$

\noindent
{\bf \uppercase\expandafter{\romannumeral 4}.}
$E^{-,i}(z)S^j(w)$. \\
For $i=j$, OPE $E^{-,i}(z)S^j(w)$ has poles at $z=q^{k+g}w$
and $z=q^{-(k+g)}w$. They come from $E^{-,i(N,1)}(z)S^{j(N,2)}(w)$
and $E^{-,i(i,1)}(z)S^{j(j+1,1)}(w)$ respectively.\\
Some terms of $E^{-,i}(z)S^j(w)$ have other poles but all these poles
cancel in pairs. Poles($z=q^{\alpha}w$) and
pairs ($E^{-,i(A)}(z)S^{j(B)}(w)$ and $E^{-,i(C)}(z)S^{j(D)}(w)$) are
$$
\begin{array}{lllllll}
 &&\alpha&(A)&(B)&(C)&(D)\\
 (\mbox{\romannumeral1})&j=i&k-N+2i+2&(i,2)&(j+1,2)&(i+2,2)&(j+2,1)\\
 &&k-N+2\ell&(\ell,1)&(\ell,2)&(\ell+1,2)&(\ell+1,1)\\
 &&&&&&i+2\leq\ell\leq N-1\\
 (\mbox{\romannumeral2})&j=i+1&k-N+2\ell-1&(\ell,1)&(\ell,2)&(\ell,2)
    &(\ell,1)\\
 &&&&&&i+1\leq\ell\leq N\\
 (\mbox{\romannumeral3})&j\leq i-1&-k-g+i-j&(j,1)&(i+1,1)&(j+1,1)
    &(i,1)\\
 &&-k-g+i-j&(j,2)&(i,1)&(j,1)&(i,2)\\
 &&-k-g+i-j&(j,2)&(i+1,1)&(j+1,1)&(i,2).
\end{array}
$$

\newpage

%
\end{document}